# Magnetic dilution on LaCrO₃

Romualdo S. Silva Jr.[a], Petrucio Barrozo [a], N. O. Moreno [a] and J. Albino Aguiar [b]

[a]Departamento de Física, Universidade Federal de Sergipe, 49100-000, São Cristóvão - SE, Brasil
[b]Departamento de Física, Universidade Federal de Pernambuco, Av. Prof. Luiz Freire, s/n, 50670-901 Recife-PE, Brasil

*Abstract — In this work, we study the effect of the magnetic dilution of $Cr^{3+}$ on the $LaCrO_3$ half doped with Al. Pure and half-doped samples were prepared by combustion method using urea as fuel. The crystal structure was investigated by X-ray diffraction. Rietveld analysis reveal a structural phase transition caused by the increase of the chemical pressure due difference of $Al^{3+}$ and $Cr^3$ ionic radii. SEM images shows that the samples exhibit a spherical formation and irregular morphologies. Magnetization measurements as a function of temperature reveal an antiferromagnetic order in both samples with a large decrease of $T_N$ (Néel temperature) with Al doping. The reduction of frustration factor and the increase of the magnetic moment of the doped sample could be attributed to the breaking of the long-range antiferromagnetic order. The magnetic hysteresis loops show a typical antiferromagnetic behavior with a slightly spin canting for the doped sample.*

## I. INTRODUCTION

The study of the substitution and doping of perovskite compounds is very important to scientific and technological development and already brought significant advances in the science [1]. It is know that the structural, electrical and magnetic properties of these compounds are strongly dependent of the level and the type of the doping [2]. Recently some studies have been done to understand and improve properties of the multiferroic materials [3, 4, 5], as well as to improve the capacity and speed of the devices such as magnetic hard disk drive [6], and magnetic sensors [7].

The LaAlO₃ has a rhombohedral structure, characterized by the rotation of the oxygen atom about the threefold axis [8]. A structural phase transition from the rhombohedral to the cubic ideal perovskite structural is observed at 813 K [9, 10]. This phase transition is pressure and temperature dependent and it is attributed to the decrease of the rotation angle [11]. The LaAlO₃ has good dielectric properties with high relative permittivity and low temperature dependence of the resonant frequency. This compound is widely used as substrate for thin films grow and it is promisors to substitute substrates in the development of the high speed device [12, 13].

The LaCrO₃ compound has been extensively studied in the last years due to the wide applicability as interconnects for solid oxide fuel cell (SOFC's), high chemical stability and good electrical properties at high temperature [14-19, 23-26]. The LaCrO₃ undergoes a structural phase transition at high temperature ranging (526 K and 413 K) from an orthorhombic to a rhombohedral structure, respectively [20], where cubic structure is suggest occur only above 1900 K [21]. This compounds presents a G-type antiferromagnetic order (AFM) below 290 K [22] and is a p-type semiconductor with activation energy $E_a = 0.19$ eV. The Néel temperature of LaCrO₃ increases with the pressure and is described by the Bloch phenomenological rule [21]. When LaCrO₃ is half-doped with Mn a competition between Cr and Mn spins is observed and this result in a reduction of the transition temperature, change of magnetic order from AFM to FM on going from LaCrO₃ to LaCr₀.₅Mn₀.₅O₃ and change on the electrical conduction mechanism [27].

Cr³⁺ ions have a d-shell semi-filled 3d³ (S=3/2) and ionic radii (R_Cr³⁺= 0,615 Å), Al³⁺ is a non-magnetic element with empty d-shell and a smaller ionic radii (R_Al³⁺= 0,535 Å) [28-31]. Partial substitution of Cr³⁺ on LaCrO₃ by Al³⁺ can induce distortion in the structure and change the magnetic properties on this leading to new interesting properties. However, until to present date we founded not find any reports in the literature about the effect of the magnetic dilution on the LaCrO₃ by Al.

## II. EXPERIMENTAL

LaCrO₃ and LaCr₀.₅Al₀.₅O₃ samples were prepared by the combustion method [32,33]. The metals salts, lanthanum nitrate (La(NO₃)₃.6H₂O), aluminum nitrate (Al(NO₃)₃.9H₂O) and chromium nitrate (Cr(NO₃)₃.9H₂O) were weighed in stoichiometric amounts. The salts are then diluted in deionized water together with urea (CH₄N₂O) as fuel. After evaporation of the excess of water and formation of the gel, the temperature was increased up to 300ºC where a combustion self-propagating reaction occurs, resulting in a fine powder whose color varies with doping. The powder was calcined at 500°C for 12 hours to burn the organics residues, after the temperature was raised to 1300°C for 12 hours where the desired phase was obtained.

The structural characterization was done by X-ray diffraction (XRD) with a Rigaku diffractometer DMAX100 using Cu-K$_{\alpha 1}$ radiation. Rietveld refinement of the diffraction patterns were performed using the General Structure Analysis System (GSAS) software. The microstructure was analyzed using JEOL microscope model JSN-5900. The magnetic properties were carried out in a VersaLab magnetometer (Quantum Design), with a magnetic field up to 3T and temperature range 50 – 350 K.

## III. RESULTS AND DISCUSSION

Figure 1 shows the results of Rietveld refinement of the powder X-ray diffraction data for LaCrO$_3$ (Fig. 1(a)) and LaCr$_{0.5}$Al$_{0.5}$O$_3$ (Fig. 1 (b)), calcined at 1300°C for 12 h. The Rietveld refinement of the patterns were performed taking as start value an orthorhombic phase, Pnma (62) indexed with the ICSD card -79344 for LaCrO$_3$ and a rhombohedra phase, belonging to R-3c (167) space group, indexed with the ICSD card 90534 of the LaAlO$_3$, for the LaCr$_{0.5}$Al$_{0.5}$O$_3$ sample. Partial substitution of Cr for Al in LaCrO$_3$ results in a structural phase transition which is accompanied by color change of the powder and confirmed by the Rietveld refinement. The parameters values obtained from the refinement are listed in Table 1.

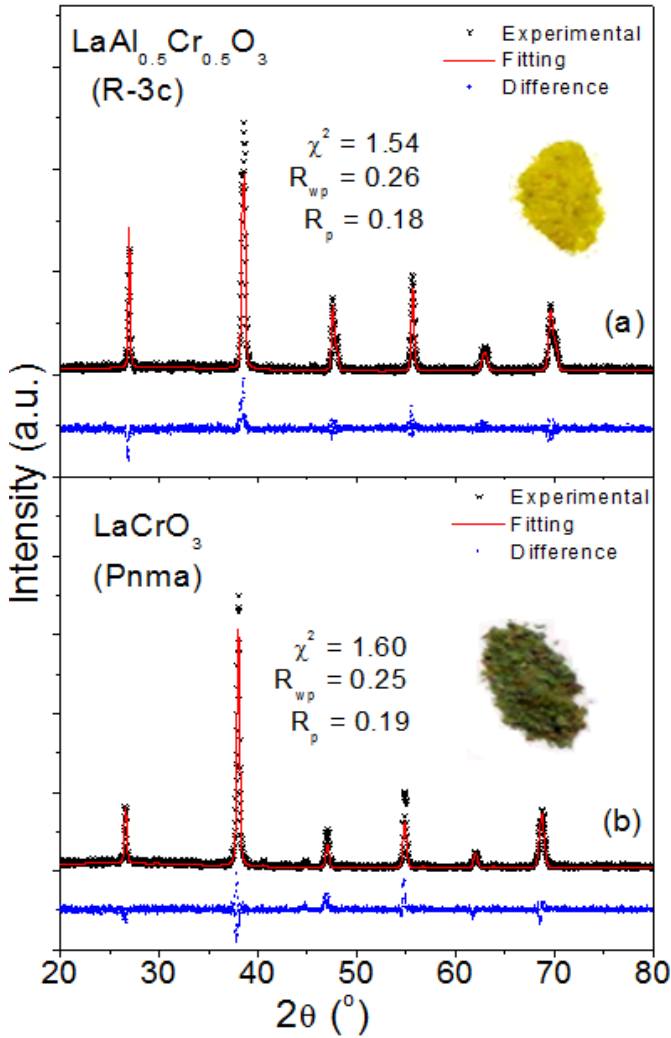

***Figure 1**: Experimental (Black dot) and, calculated (Red line) and difference (Blue dot) curves from the Rietveld refinement for samples calcined at 1300°C for 12 hours: LaCrO$_3$ (a) and LaAl$_{0.5}$Cr$_{0.5}$O$_3$ (b). The inset in (a) and (b) shows a photograph of the powder after calcination.*

**Table 1.** The Rietveld refinement parameters of X-ray diffraction data for LaCr$_{0.5}$Al$_{0.5}$O$_3$ and LaCrO$_3$ compounds collected at room temperature.

| Parameters | LaAl$_{0.5}$Cr$_{0.5}$O$_3$ | LaCrO$_3$ |
|---|---|---|
| Space group | R-3c | Pnma |
| a (Å) | 5.4379 | 5.4827 |
| b (Å) | 5.4379 | 7.7693 |
| c (Å) | 13.2034 | 5.5070 |
| α | 90 | 90 |
| β | 90 | 90 |
| γ | 120 | 90 |
| La x | 0 | 0.0170 |
| Y | 0 | 0.2500 |
| Z | 0.2500 | -0.0040 |
| Uiso | 0.0025 | 0.0003 |
| (Al,Cr) x | 0 | 0 |
| Y | 0 | 0 |
| Z | 0 | 0.5 |
| Uiso | 0.0684 | 0.0014 |
| O$_1$ x | 0.4892 | 0.3833 |
| Y | 0 | 0.2500 |
| Z | 0.25 | 0.7289 |
| Uiso | 0.0223 | 0.8000 |
| O$_2$ x | - | 0.2661 |
| Y | - | 0.4948 |
| Z | - | 0.2907 |
| Uiso | - | 0.0819 |
| Volume (Å$^3$) | 338.12 | 234.58 |
| R$_p$ (%) | 18.0 | 19.0 |
| R$_{wp}$ (%) | 26.0 | 25.0 |
| χ$^2$ | 1.54 | 1.60 |

The inset in Figure 1 shows the image of the powder obtained from the combustion reaction after heater treatment, in which can be seen the color difference between the pure (greenish powder) and half-doped sample (yellowish powder). The structural phase transition on half-doped sample can be verify by the split on the main peak located between 37°-39° as shown in Figure 2. The shift of the main peak in the doped sample to higher angles, is in accordance with Bragg's Law (2dsinθ=nλ). The minor ionic radius of Al$^{3+}$ (R$_{Al}$$^{3+}$= 0.535 Å) relative to Cr$^{3+}$ (R$_{Cr}$$^{3+}$= 0.615 Å) induce a reduction of the lattice parameter and a distortion of the CrO$_6$ octahedral causing the structural phase transition for high Al content as in half-doped sample.

The bonding angle B − O − B was calculated the Rietveld refinement, where B site is occupied by the Cr$^{3+}$ ion and for both Cr$^{3+}$ or Al$^{3+}$ in LaCrO$_3$ or LaAl$_{0.5}$Cr$_{0.5}$O$_3$, respectively. In undoped sample the B − O − B angle θ = 166.80(60)° is very close to the value found by K. Oikawa *et al.* [21]. For half-doped sample the B − O − B angle is θ = 176.48(90)° that is very close to 180° which indicate a reduction of the distortion of the octahedral.

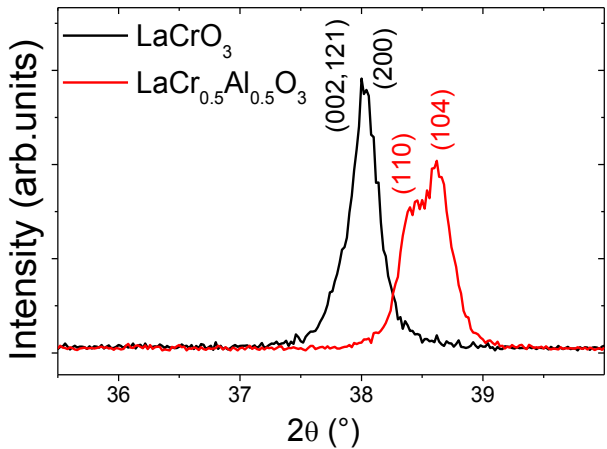

**Figure 2**: *Position of the main X-ray diffraction peak for LaCrO₃ (black line) and LaAl₀.₅Cr₀.₅O₃ (red line)*

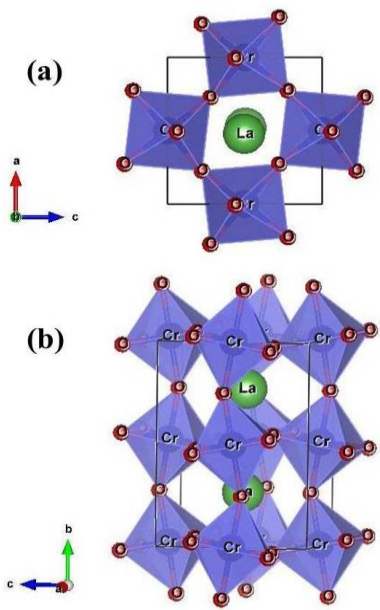

**Figure 3**: *Crystal structure obtained through the refinement of the parameters for sample LaCrO₃ projected on (a) (011) and (b) (001) plane.*

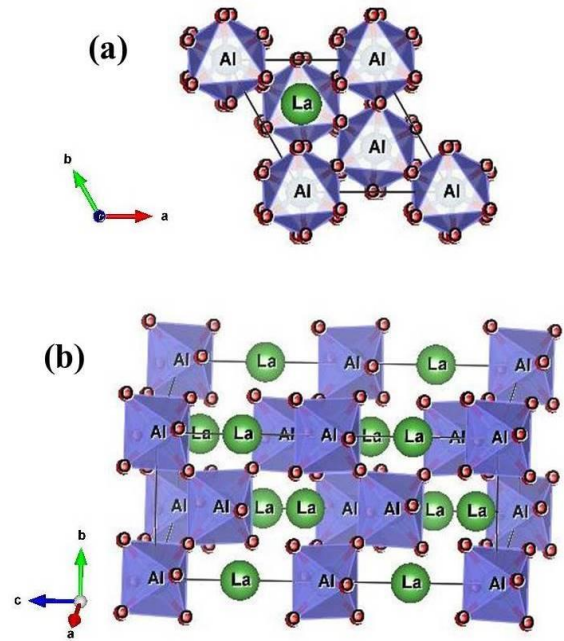

**Figure 4**: *Crystal structure obtained through the refinement of the parameters for sample LaCr₀.₅Al₀.₅O₃ projected on (a) (011) and (b) (001).*

change the magnetic order, but we can observe a strong reduction of the frustration factor (f=θ/T_N) the -9.02 at pure sample to -2.29 at half-doped sample.

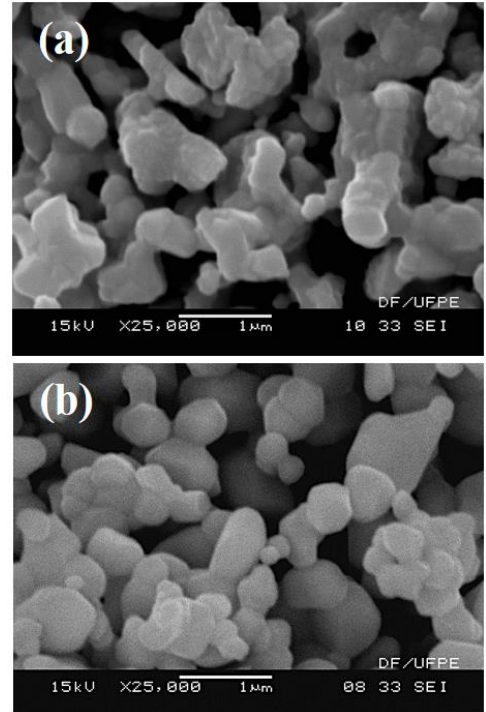

**Figure 5**:-Scanning Electron Microscopy (SEM) images for LaCrO₃ (a) and LaCr₀.₅Al₀.₅O₃ (b) samples.

Figure 3 and Figure 4 show the crystal structure obtained from the parameters summarized in Table 1 for LaCrO₃ (Figure 3a and 3b) and LaCr₀.₅Al₀.₅O₃ (Figure 4a and 4b) samples in two different orientations of the plane (011).

Figures 5(a) and 5(b) show SEM images for samples LaCrO₃ and LaCr₀.₅Al₀.₅O₃, respectively. It is observed that the samples exhibit a spherical formation and irregular morphologies with grain size close to 500 nm. With the half-doped sample with slightly smaller particles.

According Zhou et al. [28] LaCrO₃ presents an orthorhombic to hexagonal phase transition when submitted to high pressure. Nevertheless, this change does not modify the antiferromagnetic order, but it induce a shift of the spin alignment in one direction [111] of the octahedron. Our results show that Al³⁺ LaCrO₃ half-doped sample presents a similar structural phase transition with a tilt of the octahedral as show in Figure 4(b) where the octahedral have a tilt close to 45°. As related at ref. [29] the structural phase transition does not

Figure 6 shows temperature dependence of the magnetic susceptibility (χ for the pure (black) and half-doped (red) sample, respectively. In this figure, we can verify a strong decrease of the transition as well as an increase of the magnetic susceptibility can be attributed to the weakening of the long-range AFM order by substitution on Cr by Al.

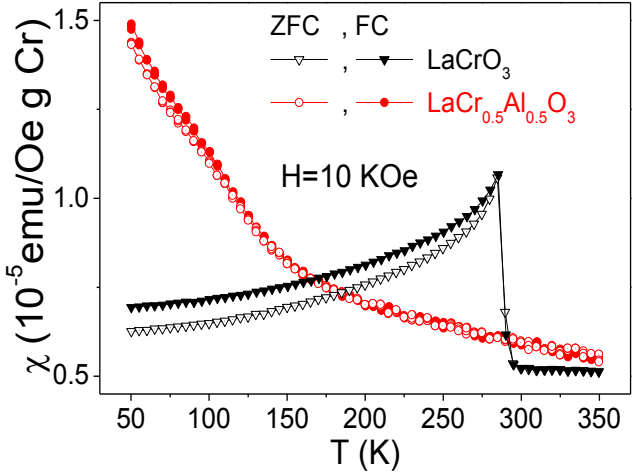

**Figure 6**: ZFC and FC magnetic susceptibility (χ) versus temperature (T) for pure sample (black) and half-doped (red) samples with H = 10 kOe.

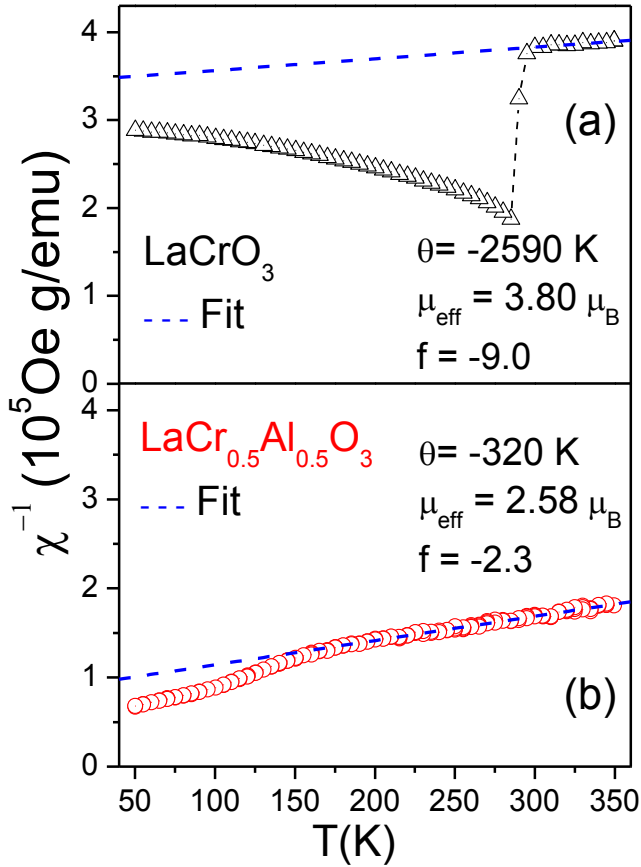

**Figure 7**: Inverse of the FC magnetic susceptibility (χ⁻¹) versus temperature (T) for pure sample (a) and half-doped (b) for H = 10 KOe.

Figure 7 shows the inverse of the FC magnetic susceptibility (χ⁻¹) as function of temperature (T) for pure (7(a)) and half-doped (7(b)) samples. The blue dashed line is

the linear fit of the FC curves in accordance with the Curie-Weiss law

$$\frac{M}{H} = \frac{C}{T - \theta} \, , \qquad (1)$$

where $C$ is the Curie constant, $\theta$ is the Weiss temperature and $T$ is the temperature.

The pure LaCrO$_3$ and half-doped sample have an antiferromagnetic order with transition temperature of 287 K and 140 K, respectively, in agreement of the values we reported previously [27]. The Curie constant and the Weiss temperature are C = 7.5x10$^{-3}$ Oe g/emu and θ = -2590 K for pure sample and C = 3. 7x10$^{-3}$ Oe g/emu and θ = -320 K for half doped sample. For LaCrO$_3$, the experimental effective magnetic moment was $\mu_{eff}$ = 3.80 $\mu_B$, which is close to the theoretically value $\mu$ = 3.87 $\mu_B$, assuming Cr$^{3+}$ with S=3/2. The half-doped sample LaCr$_{0.5}$Al$_{0.5}$O$_3$ have an experimental effective magnetic moment $\mu_{eff}$ = 2.58 $\mu_B$, that also is very close of the theoretically value $\mu$ = 2.74 $\mu_B$, assuming Cr$^{3+}$ with S=3/2 and Al$^{3+}$ with S=0.

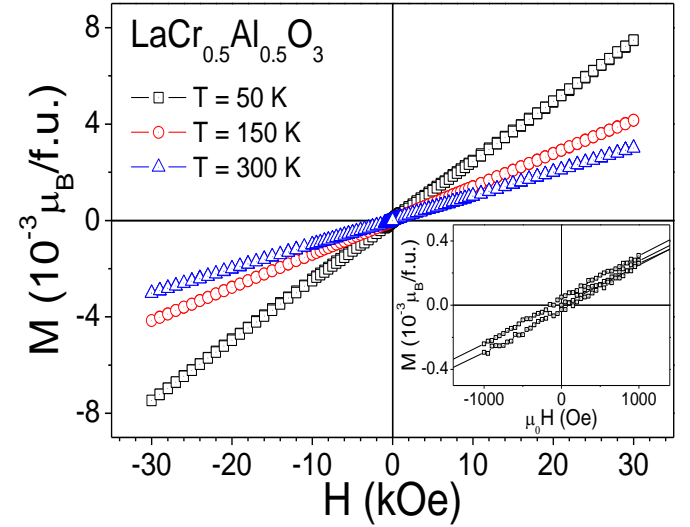

**Figure 8**: Magnetization as a function of applied field for half-doped sample LaCr$_{0.5}$Al$_{0.5}$O$_3$ at different temperatures T = 50 K (black), 150 K (red) and 300 K (blue). The inset shows the low-field behavior at 50.

Magnetization as a function of applied magnetic field curves for half-doped sample at 50 K (black), 150 K (red) and 300 K (blue) are shown in Fig. 8. We can observe a typical paramagnetic behavior at 150 K and 300 K. At 50 K it is observed a small opening of the magnetic isotherm which can be attributed to the canting of the antiferromagnetically ordered Cr$^{3+}$ spins.

Figure 9 shows the magnetization curves as a function of magnetic field for pure and half-doped sample at 300 K. We can observed that the pure sample with non-diluted magnetic moment have smaller magnetization compared with diluted sample, this could be attributed to the break of the long range antiferromagnetic order that result in an uncompensated magnetic moment and increase the magnetization. We noted that doping of Al into LaCrO$_3$ enhances the magnitude of magnetization at 300 K.

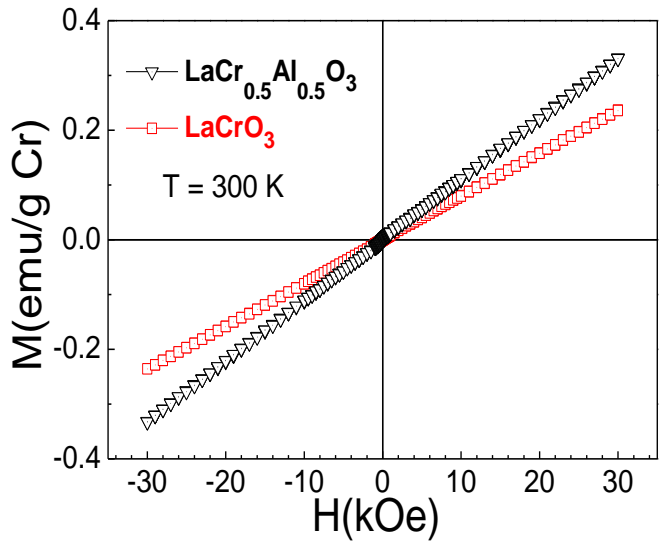

**_Figure 9_**: Magnetization as a function of applied field for LaCrO$_3$ (black) and LaCr$_{0.5}$Al$_{0.5}$O$_3$ (red) at 300 K.

## IV. CONCLUSION

LaCrO$_3$ pure and Al half-doped samples prepared by combustion method were successful obtained. The pristine sample (LaCrO$_3$) presents an orthorhombic structure with space group *Pnma (62)* and the Al half doped sample LaAl$_{0.5}$Cr0.5 a romboedral structure with space group *R-3c* for addition of Al in LaCrO$_3$ induces a shift of the peaks of the x-ray diffraction to higher angles, due to the difference of the ionic radius of Al$^{3+}$ and Cr$^{3+}$. SEM micrographs reveal that the samples exhibit a spherical formation with size close to 500 nm and irregular morphologies. The magnetic measurements reveal that the Al half-doped samples induce a break of the long-range antiferromagnetic interaction resulting in an uncompensated magnetic moment that is responsible for the increase of the magnetic moment, of the half-doped sample compared to that of the pristine one. In addition, the doped sample has a large suppression of the magnetic transition temperature. A small irreversibility between the ZFC and FC modes below 140 K and a small opening of the hysteresis loop at 50 K is observed at half-doped sample. A color change also is observed changing from dark green (LaCrO$_3$) to yellow (LaAl$_{0.5}$Cr$_{0.5}$O$_3$).

## V. ACKNOWLEDGMENTS

This work was supported by the Brazilian science agencies CAPES (grant PNPD 2498/2011), CNPq (grants 443458/2014-6, 307552/2012-8 and 141911/2012-3) and FACEPE (grant APQ-0589/1.05-08).